\documentstyle[12pt]{article}

\newcommand{\be}{\begin{equation}}
\newcommand{\ee}{\end{equation}}
\newcommand{\bea}{\begin{eqnarray}}
\newcommand{\eea}{\end{eqnarray}}

\makeatletter

\def\eqnarray{\stepcounter{equation}\let\@currentlabel=\theequation
\global\@eqnswtrue
\global\@eqcnt\z@\tabskip\@centering\let\\=\@eqncr
$$\halign to \displaywidth\bgroup\@eqnsel\hskip\@centering
$\displaystyle\tabskip\z@{##}$&\global\@eqcnt\@ne
\hfil$\displaystyle{\hbox{}##\hbox{}}$\hfil
&\global\@eqcnt\tw@ $\displaystyle\tabskip\z@
{##}$\hfil\tabskip\@centering&\llap{##}\tabskip\z@\cr}
\@addtoreset{equation}{section}
\def\theequation{\thesection.\arabic{equation}}
\makeatother
\begin{document}
\begin{titlepage}
\date{June 1994,\\ DTP/94/19}
\title{A Fresh Look at Generalized Veneziano Amplitudes}

\author
{David B. Fairlie\\
{ \it University of Durham, South Road,}\\
Durham, DH1 3LE, UK\\
\phantom{}\\
Jean Nuyts\\
{ \it University of Mons-Hainaut,}\\
7000, Mons, BELGIUM}
\maketitle
\begin{abstract}
The dual resonance model, which was a precursor of string theory was based
upon the idea that two-particle scattering amplitudes should be expressible
equivalently as a sum of contributions of an infinite number of
$s$ channel poles each corresponding to
a finite number of particles with definite spin, or as a similar sum of $t$
 channel poles. The famous example of Veneziano \cite{ven}
satisfies all these requirements, and is additionally ghost free.We recall
other trajectories which provide solutions to the
duality constraints, e.g.
 the general Mobi\"us trajectories and the
logarithmic trajectories, which were thought to be lacking this
last feature. We however demonstrate, partly empirically, the
existence of a regime within a particular deformation of the
Veneziano amplitude for logarithmic trajectories for which the
amplitude remains ghost free.

\end{abstract}
\thispagestyle{empty}
\end{titlepage}
\section{Introduction.}
The origin of String Theory
as an interpretation of the celebrated dual 4-point
amplitude of Veneziano \cite{ven} seems to us in danger of being forgotten.
In this article we re-examine the ideas behind the
concept of dual amplitudes in
the hope of
opening further avenues for the exploration of new String Theories.
This concept,
as applied to 4-point tree amplitudes, required that the amplitude could
alternatively be expressed either as a sum of an infinite number
of $s$ channel poles at positions
$s=\lambda_n$, with residues given by polynomials of maximal degree $n$ in the
dual variable $t$, or as a sum of $t$ channel poles with polynomial residues
in $s$. There was a further constraint which was technically very difficult to
implement; the partial wave decomposition of the residues into appropriate
angular functions (Legendre Polynomials in the case of 3+1 dimensional
space-time) should have only positive coefficients, otherwise the corresponding
particles exchanged would be ghosts.

The original solution of Veneziano to the duality constraints was to take a
linear (mass squared) spectrum, i.e. a linearly rising Regge trajectory. He
proposed an amplitude of the well known Euler $\beta$-function form;
\be
\prod_{n=0}^{\infty}\frac{s+t-\lambda_n-\lambda_0}{(s-\lambda_n)(t-\lambda_n)}\
 \ ,
\label{venez}
\ee
which has all the required properties, including freedom from ghosts for the
choice
\be
\lambda_n=\alpha^{-1}n+\lambda_0,
\ee
a linear relation with slope $\alpha$ and intercept $\alpha_0=
-\alpha\lambda_0$,
provided $\alpha\lambda_0$ lies between -1 and 0. By rescaling $\alpha$ can be
 put to 1.

This construction rapidly led to the development of string theory in the late
sixties and early seventies, together with all the machinery of a Fock space
representation in terms of an infinite number of creation and annihilation
operators \cite{fub}. Another solution
to the duality constraint, resulting in the so-called logarithmic trajectories
was found independently by several authors \cite{bc,cn,go,ma,ro}. That of
 \cite{cn} is the
unique solution for the dual 4-point amplitude when the residues are
polynomials of finite extent from $t^{n-Q}$ (or obviously from $t^0$  when
$n<Q$) to $t^n$ where $Q$ is a fixed value independent of $n$. These
logarithmic
trajectories are of the form
\be
\lambda_n=a\sigma^n,
\label{logarithm}
\ee
a geometric series. A simple scaling and shift for $s,t$,
redefining\hfil\break
$s\mapsto a(\sigma-1)s+1,\ t\mapsto a(\sigma-1)t+1$ yields a pole spectrum
\be
\lambda_n=l(n)
\ee
with
\be
l(n)=\frac{\sigma^n-1}{\sigma-1},\quad {\rm i.e.}\quad
0,1,1+\sigma,1+\sigma+\sigma^2\ldots,
\label{spectrum}
\ee
showing that in the limit $\sigma\rightarrow 1$, the linear mass spectrum is
recovered.
In the regime $\sigma >1$ for which these amplitudes were thought applicable,
they possess ghosts and this is perhaps the principal reason for which their
study was abandoned. In this context it is amusing to note that although Baker
and Coon and
Cremmer and Nuyts discovered the same trajectory, these pairs of authors had
different 4-point functions, neither of which was ghost free\ ! It is not
surprising that a satisfactory operator realisation connecting in the
appropriate limit with the usual string
operator formalism has not been found,
despite some attempts \cite{coon3,masud}.
One has too few sets of oscillators, the other too many\ !

We find empirical evidence that in the region $0\leq\sigma\leq 1$ the amplitude
proposed by Baker and Coon when adjusted to correspond with the Veneziano
 amplitude in the limit  $\sigma\rightarrow 1$ is indeed ghost free. In this
 region the sequence
$\lambda_n $ tends to a limit point, and this would give rise to an essential
singularity in this approximation to the scattering amplitude. In terms of the
 thinking in the 1970's
with an interpretation of the amplitude as a hadronic amplitude, this was
unacceptable, and Baker and Coon
chose the range of their parameter such that $\sigma>1$, with consequent ghost
behaviour.
However, in view of much recent discussion reviving the old idea of a limiting
distance
\cite{asp} and the modern re-interpretation of strings
as describing elementary particles
with a scale appropriate to the Planck mass, it may not be too fanciful to
suggest that
these models with a spectrum with an accumulation point indicate the
possibility of
a limit to the energy in an elementary scattering process. Similar ideas have
arisen in
speculations concerning deformations of the Poincar\'e group.
It should also be borne in mind that many perfectly respectable physical
systems, for example, the hydrogen atom and positronium, possess an infinite
number of bound states with an accumulation point.

In the course of this investigation we rediscovered a more general solution to
the duality constraint than that of the logarithmic trajectories \cite{bc}.
This corresponds to
a spectrum of the form
\be
\lambda_n=\frac{a\sigma^n+b}{c\sigma^n+d},\ \ \ \ a,b,c,d,\sigma\ \ {\rm
 constant},
\ee
 which we call a M\"obius trajectory. To our surprise, we found that it had
 already
been found by Coon in his very first paper on the subject
\cite{coon2}\ ! The
reason that it was dropped from subsequent papers, we suppose, is that if all
pole positions are positive, then, whatever the value of $\sigma$ the model has
an accumulation point, except for the linear and exponential
cases (\ref{logarithm}).
While the 4-point amplitude is viable, we have not
succeeded in constructing a candidate 5-point amplitude with polynomial
residues of the correct degree except in the logarithmic ($c=0$) case.

The outline of this paper is as follows. In section 2 we justify the M\"obius
trajectory solution \cite{coon2} in our own way. In section 3 we extend the
analysis to the 5-point function,
and show that the Baker and Coon ans\"atz satisfies recurrence
relations similar
to those proposed for the linear spectrum by Bardakci and Ruegg \cite{br}.
Vice-versa these recurrence relations are sufficient to determine the 5-point
amplitude up to an overall multiplication constant.

The question of the absence of ghosts which is equivalent to the
positivity of the coefficients of the expansion of the
residues in terms of orthogonal Legendre polynomials
is addressed in the final section. Since this is an extremely
difficult analytic problem and since the result was proved for the Veneziano
amplitude only after operator factorisation and the no-ghost theorem
\cite{god,brow} we have had to recourse to computer verification using
REDUCE. Thus our results on the domain of parameters
for which our amplitude is
ghost free remain conjectural, and await an operator realisation. On the
 other
hand, our results are sufficiently positive as to encourage belief in the
existence of such a representation. Something along the lines of a q-deformed
operator realisation could be expected \cite{fai}.

\section{Outline of the problem. The Generalized Linear Case.}
 We suppose that the dual 4-point amplitude for four
incoming particles of momenta $p_i,\ i=1,2,3,4,\ \sum p_i=0$
can be written in the
following form symmetrical in $s=(p_1+p_2)^2$ and $t=(p_2+p_3)^2$
which generalizes in an obvious
way the usual solution (\ref{venez})
\be
A(s,t)=N\prod_{r=0}^\infty \frac{\gamma_r
(s+t)-\alpha_r-\beta_rst}{(s-\lambda_r)(t-\lambda_r) }\ \ ,
\label{2.1}
\ee
where $N$ is a suitably chosen, possibly infinite
normalisation constant,
the $\lambda_r$ are the positions of the poles
both in the $s$ and $t$
channels, while $\hat\alpha_r=\alpha_r/\gamma_r$ and
$\hat\beta_r=\beta_r/\gamma_r$ are
infinite sets of parameters
which will be restricted by the physical requirements outlined
in the introduction. The set of parameters $\gamma_r$ which can be
absorbed in the normalisation constant $N$ is introduced for later
convenience. We call this case the generalized linear
case as, when evaluated for one variable at one of its poles,
every term in the
numerator is of first degree in the other variable. The
numerator in the right-hand side of (\ref{2.1}) is obviously the most
general product of polynomials symmetric under the
exchange of $s$ and $t$ with this property.
In agreement with the ideas expressed
in the introduction we will now
require that when evaluated at a pole $\lambda_n$
in the $s$-variable the
remaining $t$-dependence of $A$ is restricted to a
polynomial of degree
$n$ at most. This implies that the $t$-poles at all the values
$\lambda_r$ should be killed by zeros
arising in the numerators
at values
\be
t_m={\hat\alpha_m-\lambda_n \over 1-\hat\beta_m\lambda_n}.
\label{2.2}
\ee
We shall now suppose that this is done in a sequential way.
The $n$
first terms in the product in the numerator survive.
The next terms are
suppressed sequentially in the numerator and denominator.
Hence we require
\be
t_{n+r}=\lambda_r
\label{2.3}
\ee
for all $r\geq 0$ and $n\geq 0$. These relations become
\be
E(n,r)\equiv
\hat\alpha_{n+r}-\lambda_n-\lambda_r+\hat\beta_{r+n}\lambda_r\lambda_n=0.
\label{2.4}
\ee
At first sight the problem appears overdetermined but
fortunately many relations are redundant. Let us note first that
the equations are
symmetrical in $n$ and $r$ so that we can restrict ourselves to
$n \geq r$.
The sequential solution of this (doubly infinite) set of
equations can be obtained easily as follows. Indeed let us
take a set of $n$
and $r$ values such that $n+r$ is fixed. The first four values
$n+r=0,1,2,3$ are special and have to be treated separately.
Then
starting at $n+r=4$ we obtain the general
regime.
\begin{itemize}
\item[0)] For $n+r=0$ we only have one
relation between $\lambda_0$, $\hat\beta_0$ and $\hat\alpha_0$.
Hence two of them
(say $\lambda_0$ and $\hat\alpha_0$) can be considered
as free parameters.
\item[1)] For $n+r=1$ there
is again only one relation for $n=1,\ r=0$ between
$\hat\beta_1$, $\hat\alpha_1$,
$\lambda_0$ and $\lambda_1$.
Hence there are two new free parameters (say
$\lambda_1$ and $\hat\alpha_1$) .
\item[2)] For $n+r=2$ there are two relations $n=2,\ r=0$
and $n=1,r=1$ which
allow the determination of $\hat\beta_2$ and $\hat\alpha_2$ with one new free
parameter $\lambda_2$.
\item[3)] For $n+r=3$ there are again two relations $n=3,\ r=0$ and
$n=2,\ r=1$which
allow the determination of $\hat\beta_3$ and $\hat\alpha_3$
with one new free
parameter $\lambda_3$.
\item[4)] For $n+r=p\geq 4$ there are three relations or more. Three of
them allow the determination of $\hat\beta_{p}$, $\hat\alpha_{p}$ and
$\lambda_{p}$ in terms of $\lambda$'s of lower values. These
determinations are sequential. We write them down explicitly~:
\end{itemize}
\begin{eqnarray}
&{}&\lambda_p=\nonumber\\
&{}&\frac{\lambda_{p-2}\lambda_{p-1}\lambda_2
-\lambda_{p-2}\lambda_{p-1}\lambda_1
 +\lambda_{p-2}\lambda_{2}\lambda_{1}
-\lambda_{p-2}\lambda_{2}\lambda_{0}
-\lambda_{p-1}\lambda_{2}\lambda_{1}
 + \lambda_{p-1}\lambda_{1}\lambda_{0}}
{ \lambda_{p-2}\lambda_{2}
 -\lambda_{p-2}\lambda_{0}
 -\lambda_{p-1}\lambda_{1}
 + \lambda_{p-1}\lambda_{0}
 -\lambda_{2}\lambda_{0}
 + \lambda_{1}\lambda_{0}}\nonumber\\
&{}&\hat\beta_{p} =
\frac{\lambda_{p-2} -\lambda_{p-1}
+ \lambda_{2} -\lambda_{1}}
 {\lambda_{p-2}\lambda_{2}
-\lambda_{p-1}\lambda_{1}}\label{2.5}\\
&{}&\hat\alpha_{p}=
\frac{\lambda_{p-2}\lambda_{p-1}\lambda_{2}
 -\lambda_{p-2}\lambda_{p-1}\lambda_{1}
 + \lambda_{p-2}\lambda_{2}\lambda_{1}
 -\lambda_{p-1}\lambda_{2}\lambda_{1}}
 { \lambda_{p-2}\lambda_{2}
-\lambda_{p-1}\lambda_{1}}.\nonumber
\end{eqnarray}
\begin{itemize}
\item[] The remaining relations apart from the first three for
$n+r$ fixed are then automatically satisfied.
The easiest way to see this is from the general
solution to (\ref{2.5}) which we shall obtain.
\end{itemize}

Apart from the trajectory, which we shall discuss
next, the amplitude
(up to the arbitrary normalisation factor $N$) depends upon two
further parameters which may be taken to be
$\hat\alpha_0$ and $\hat\alpha_1$.
\par The Regge trajectory,
which is the important physical
quantity, has four arbitrary parameters
$\lambda_0,\lambda_1,\lambda_2$ and $\lambda_3$ with the higher
poles $\lambda_p,p\geq 4$ given by (\ref{2.5}).
After choosing $\gamma_r$ suitably, we find that the general solution can be
written as
\bea\label{trajectory}
\lambda_r&=\frac{a\sigma^r+b}{ c\sigma^r+d}.
 \phantom{ \ \ \ \ \ {\rm{for}} \ r>1}\\
\alpha_r&=\frac{a^2\sigma^r-b^2 }{ c\sigma^r+d}
\ \ \ \ \ {\rm{for}} \ r>1.\\
\beta_r&=\frac{c^2\sigma^r-d^2 }{ c\sigma^r+d}
\ \ \ \ \ {\rm{for}} \ r>1. \\
\gamma_r&=\frac{ac\sigma^r-bd}{ c\sigma^r+d}
\ \ \ \ \ {\rm{for}} \ r>1.
\eea
we refer to such a spectrum as a M\"obius trajectory.

To be complete let us quote the two remaining equations which are associated
with the two first values of $n+r$
\bea
&\hat\alpha_0-2\lambda_0+\hat\beta_0\lambda_0^2=0 \\
&\hat\alpha_1-\lambda_0-\lambda_1+\hat\beta_1\lambda_0\lambda_1=0.
\eea
However to make everything nicely continuous in the
discrete variable $r$ we will choose the two free remaining
parameters $\hat\alpha_0$ and $\hat\alpha_1$ in such a way that (2.7),(2.8)
and (2.9) are valid even for $r=0$ and $r=1$.
\par The four independent (complex) parameters, three in the
$SL(2,C)$ M\"obius matrix $M$
\be
M=\left(\begin{array}{cc} a & b\\
c &d\end{array}\right),\ \ \ \det M=ad-bc=1
\label{2.7}
\ee
and $\sigma$,
are then determined in terms of the first four pole positions
(in principle complex also)
$\lambda_0,\lambda_1,\lambda_2$ and $\lambda_3$, by solving three linear
and one quadratic equation.
The parameter $\sigma$ is determined as any solution
of the equation
\be
\sigma^2 +\sigma(z+1)+1=0
\label{2.11}
\ee
where $z$ is the cross ratio
\be
z=\frac{(\lambda_3-\lambda_0)(\lambda_1-\lambda_2)}
{(\lambda_1-\lambda_0)(\lambda_3-\lambda_2)}\ \ .
\ee
Regarding the equations (\ref{2.5}) for fixed $n+r$
as linear equations for $\hat\alpha(n+r)$ and $ \hat\beta(n+r)$
the solution (2.6) can be verified to ensure that only two
of these equations are linearly independent, and thus that
the solution is fully consistent.
\par The invariance of the manifold of the trajectories is
essentially $SL(2,C)\otimes GL(1,C)$ where $GL(1,C)$ corresponds to the
multiplication of $\sigma$ by an arbitrary complex number. The
trajectories themselves are invariant under the $Z_2$ transformation
which interchanges $\sigma$ and $1/\sigma$ while also interchanging
$a\leftrightarrow b$ and $c\leftrightarrow d$.
\par Regarding purely real trajectories, since $\lambda_0$ and $\lambda_1$
can be put to $0$ and $1$
respectively by a translation (fixing the intercept) and a dilation
(fixing the energy scale), the two last
parameters $\lambda_2$ and $\lambda_3$ are
those which fix the shape of the trajectory.
Let us remark here that this shape is very
sensitive to the values of the remaining physical parameters
$\lambda_2$ and $\lambda_3$. Some trajectories saturate to a finite value
when $r$
goes to $\infty$, others go to $\infty$. There are other trajectories
which behave as tangents but should probably be excluded physically since
they involve tachyons. Allowing some of the parameters to become
complex can lead, for example, to trajectories whose few first
members are real and hence stable while the higher ones are
complex and hence intrinsically unstable.
\par Remarkably, all three expressions
$\lambda_r,\ \hat\alpha_r$ and $ \hat\beta_r^{-1}$
all satisfy the same M\"obius type recurrence relation
\be
\lambda_{r+1}=
\ \frac{v\lambda_r+w}{
x\lambda_r+y}\ {\buildrel \rm def \over =}\ \lambda_r^{[1]},
\label{2.12a}
\ee
with
\bea
&v=bc-ad\sigma \nonumber\\
&w=ab(\sigma-1)\nonumber\\
&x=cd(1-\sigma) \nonumber \\
&y=bc\sigma-ad.\nonumber
\label{2.12b}
\eea
The perceptive reader may wonder how it comes about
that the general solution of
a sequence $\lambda_n$ which satisfies a second order recurrence relation
(\ref{2.5}) also satisfies a first order relation (\ref{2.12a}).
The reason is
that (\ref{2.5}) also encodes the boundary values to be satisfied by
$\lambda_n$.

Armed with these relations,
it is possible to show that the amplitude $A(s,t)$
defined in equation (\ref{2.1}) satisfies the functional equation
\be
A(s,t)=
\frac{(\gamma_0(s+t)-\alpha_0-\beta_0st)}{(s-\lambda_0)}A(s^{[-1]},t)
\label{2.13}
\ee
and
\be
s^{[-1]}=\frac{-ys+w}{xs-v}.
\label{2.14}
\ee
As expected, the transformation between
$s$ and $s^{[-1]}$ is the inverse of the
M\"obius transformation (\ref{2.12a}).

When the trajectory is restricted to be purely logarithmic, say
(\ref{logarithm}) with $a=1$, defining $\tau=1/\sigma$ and
noting that $s^{[-1]}=\tau s$, one
finds a recurrence relation
\be
A(s,t)=
sA(s,\tau t)+A(\tau s,t),
\ee
which may be re-cast in the form of a $q$-derivative-difference
equation
\be
D_{s}A(s,t)=\frac{1}{1-\tau}A(s,{\tau}t),
\label{rec4}
\ee
where
\be
D_{s}A(s,t)=
\frac{A(\tau s,t)-A(s,t)}{s(\tau-1)}.
\ee
Equations of the form (\ref{rec4}) determine the 4-point
function up to a constant and will be used more systematically
for the 5-point function.

\section{The five point function.}

Let as usual
$s_{ij}=(p_i+p_j)^2,i,j=1,\ldots,5$ denote the
generalized Mandelstam variables in terms of the incoming
momenta of five scalar particles. Consider the usual planar
tree diagrams
and the particular subset of variables
$\displaystyle{s_k\buildrel \rm def \over = s_{k,k+1}}$
where poles
at positions $\lambda_m$
due to the tree diagrams can occur.
These poles must obey the duality, factorization
and symmetry restrictions~:
\begin {itemize}
\item[a)] The residue of any pole in any of
the $s_A$ variable, say $s_{1}$
should have no pole in the neighbouring variables i.e.
no poles in the
variables $s_{5}$ and $s_{2}$. Double poles can occur only in
non-neighbouring variables.
\item[b)] The residues of the allowed double poles
should factorize in the usual way.
\item[c)]The amplitude should have the $D_5$ symmetry of the pentagon.
It consists in the cyclic $C_5$ symmetry permuting the five legs
of the dual diagram. Under this symmetry the five variables
$s_{1},s_{2},s_{3},s_{4},s_{5}$ are rotated cyclically.
The extra symmetry needed to define $D_5$ by closure under products
is generated by, say, the mirror symmetry
$Z_2^{(1)}$ generated by the simultaneous
interchange of the variables
$s_{2}\leftrightarrow s_{5}$ and
$s_{3}\leftrightarrow s_{4}$ leaving $s_{1}$ fixed.
\end{itemize}

The Baker Coon 5-point function, which
is again more conveniently expressed in terms of $\tau =1/\sigma$,
is constructed as a multiple sum
\bea
&&A(s_{1},s_{2},s_{3},s_{4},s_{5})=\nonumber\\
&&\sum_{{\rm all}\ n_{i}=0}^{\infty}
\frac{s_{1}^{n_{1}}}{f_{n_{1}}}\tau^{n_{1}n_{2}}
\frac{s_{2}^{n_{2}}}{f_{n_{2}}}\tau^{n_{2}n_{3}}
\frac{s_{3}^{n_{3}}}{f_{n_{3}}}\tau^{n_{3}n_{4}}
\frac{s_{4}^{n_{4}}}{f_{n_{4}}}\tau^{n_{4}n_{5}}
\frac{s_{5}^{n_{5}}}{f_{n_{5}}}\tau^{n_{5}n_{1}}
\label{bake}
\eea
which is obviously $D_5$ invariant and where
\be
f_{n} = (1-{\tau})(1-{\tau^2})\cdots(1-{\tau^n}).
\ee
They show that 4 of the 5 summations can be performed formally
yielding
\bea
&&A(s_{1},s_{2},s_{3},s_{4},s_{5})=\nonumber\\
&&\sum_{n_{4}=0}^{\infty}
\frac{G({s_{5}s_{1}\tau^{n_{4}}})}{G({s_{5}}{\tau^{n_{4}}})G(s_{1})}
\frac{G({s_{2}s_{3}}{\tau^{n_{4}}})}{G(s_{2})G({s_{3}\tau^{n_{4}}})}
\prod_{r=1}^{n_{4}}
\frac{s_{1}s_{2}-s_{4}\tau^r}{1-\tau^r}.
\label{co}
\eea
Here $G(z)$ denotes the function
\be
G(z)= \prod_{r=0}^{\infty}(1-z\tau^r).
\label{product}
\ee
The basic result used is the Euler sum
\be
\sum_{n=0}^{\infty}\frac{z^n}{\prod_{r=1}^{r=n}(\tau^r-1)}=
\prod_{r=0}^{\infty}
\frac{1}{(1-z\tau^{r})} =E_\tau(\frac{z\tau}{1-\tau}).
\ee
Here $E_{\tau}(z)$ denotes the Jackson \cite{jack} form of the
$q$-deformed exponential for
$q=\tau$. This is meaningful, i.e. the series converges uniformly and
absolutely for all finite $z$ if $\tau\leq1$, but for $\tau>1$ ($|\sigma|<1$)
the series converges uniformly and absolutely for $|z|<1$ and diverges
otherwise. This is the range for which the amplitudes are found to be ghost
free.

In this representation it is easy to see that their amplitude has no
simultaneous poles in adjacent variables, but does admit double poles in
$s_{3}$ and $s_{5}$ at $s_{3}=\sigma^{k_3}$ and $s_{5}=\sigma^{k_5}$ with
residue a polynomial in the variables $s_{1},\ s_{2},\ s_{4}$
built from monomials of the form
\be
s_1^{k_5-\beta}s_2^{k_3-\beta}s_4^{\beta}\ ,\ \ 0\leq\beta\leq\min(k_3,k_5)
\label{residue5}
\ee
as can easily be seen from (\ref{co}). This is in agreement with the
standard Feynman tree amplitude.

In their original derivation of the 5-point function for linear trajectories
\cite{br}, Bardakci and Ruegg developed an integral
representation for their amplitude, by analogy
with that for the 4-point, which exhibits the correct double poles.
They went on on to develop an
alternative approach to the 5-point based
upon functional recursion relations
which it must satisfy and which are derived
by partial integration of their 5-point representation.
We have derived just such a recurrence
relation in section 2 for our 4-point function and now display analogous
recurrence relationships which are satisfied by Baker and Coon's
dual 5-point function. After performing permutations of the variables
one obtains altogether five equations. A typical member is as
follows;

\bea
A(s_{1},s_{2},s_{3},s_{4},s_{5})&=&
s_{5}A({\tau}{s_{1}},s_{2},s_{3},{{\tau}s_{4}},s_{5})\nonumber\\
&+&A(s_{1},s_{2},s_{3},s_{4},{\tau}{s_{5}}).
\eea
These equations are derived from the representation in the form of equation
(\ref{bake}). They are analogous to the recurrence equation
(\ref{rec4}) we have found for the 4-point function and may be
re-cast in the form of a $q$-derivative-difference
equation
\be
D_{s_{5}}A(s_{1},s_{2},s_{3},s_{4},s_{5})=\frac{1}{1-\tau}
A({\tau}{s_{1}},s_{2},s_{3},{\tau}{s_{4}},s_{5}).
\label{rec}
\ee

These equations can be used alternatively to define a  5-point
amplitude.
Indeed, from these equations it is easy to prove the following
relations
\bea
&\frac{\partial^{m_1+m_2+m_3+m_4+m_5}}{(\partial s_1)^{m_1}(\partial
s_2)^{m_2}(\partial s_3)^{m_3}
(\partial s_4)^{m_4}(\partial s_5)^{m_5}}A_{s_1=s_2=s_3=s_4=s_5=0} \nonumber\\
&\ \ \ =m_1\frac{\tau^{m_2+m_5}}{ 1-\tau^{m_1}}
\frac{\partial^{m_1+m_2+m_3+m_4+m_5-1}}{(\partial s_1)^{m_1-1}(\partial
s_2)^{m_2}(\partial s_3)^{m_3}
(\partial s_4)^{m_4}(\partial s_5)^{m_5}}A_{s_1=s_2=s_3=s_4=s_5=0}.\nonumber\\
\label{3.2a}
\eea

 From the basic equations (\ref{rec}) numerous other identities
can be derived. The following ones are particularly interesting in the light
they cast upon the pole structure of the amplitude
\bea
A(s_{1},\tau s_{2},s_{3},s_{4},\tau s_{5})&=&\frac{1}{
s_{1}-1}\times\nonumber\\
\Bigl(
(s_{2}-1)A(\tau s_{1},s_{2},\tau s_{3},s_{4},s_{5})
&+&(s_{5}-s_{3})A(\tau s_{1},\tau s_{2},s_{3},\tau s_{4},s_{5}).
\Bigr)
\label{recc}
\eea

These equations may be used to determine the values of the expansion
coefficients in a multiple power series in $s_{i}$ for the 5-point
amplitude. It is easy to prove from
the recurrence relations satisfied by these
coefficients that (\ref{3.2a})
and its cyclic partners determine this amplitude
uniquely, up to a multiplicative constant, as that of Baker and Coon.

The substitution of the
M\"obius transformed
values of $s_i^{[M]}$in the amplitude $A(s_i)$ of Coon and Baker (\ref{bake})
gives rise to a M\"obius amplitude $A^{[M]}(s_i)$
\be
A^{[M]}(s_i)=A(s_i^{[M]})
\ee
where, see (\ref{trajectory}),
\be
s_i^{[M]}=\frac{ds_i-b}{-cs_i+a}
\ee
which obviously  has poles at the M\"obius spectrum (\ref{2.12a})
with no simultaneous neighbouring poles.
Unfortunately the residues of these poles are of too
high degree to permit an interpretation in terms of the exchange of particles
with spin no greater than that associated with the
corresponding pole (\ref{residue5}). The corresponding identities
(\ref{recc}) are written
\bea
(s_{1}-\lambda_0)&A&(s_{1},s_{2}^{[-1]},s_{3},s_{4},s_{5}^{[-1]})-
(s_{2}-\lambda_0)A(s_{1}^{[-1]},s_{2},s_{3}^{[-1]},s_{4},s_{5})\nonumber\\
&=&(s_{5}-s_{3})A(s_{1}^{[-1]},s_{2}^{[-1]},s_{3},s_{4}^{[-1]},s_{5})
\label{recm}
\eea
using the inverse M\"obius transformed variables
$s_i^{[-1]}$ defined previously (\ref{2.14}).

\section{Ghost Freedom}

One of the most seredipitious properties of the Veneziano amplitudes is
the absence of ghosts, i.e. the absence of contributions
from intermediate states of integral spin with
negative coefficients in dimensions $\leq 26$, when the ground state mass
squared lies between -1 and 0. (The former limit corresponds to a ground state
tachyon, for which the theory admits an operator factorisation.) Indeed the
only analytic way known to us to demonstrate freedom from ghosts is via the
operator formalism \cite{god,brow}.

It would be remarkable if this property were not to continue to hold when the
Veneziano amplitude is slightly deformed. As there is no obvious way to test
this analytically,
the only way to proceed is to test for a range of cases using
an algebraic computation programme (in our case REDUCE).

We have taken the
basic Baker and Coon amplitude and peformed a rescaling as explained in the
introduction to make the $\sigma=1$ limit coincide exactly with the Veneziano
amplitude with slope 1 and a shift in such a way that the scalar external
particle be of (rescaled) mass $m$ with $-1\leq m^2\leq 0$
\be
\lambda_n=\frac{\sigma^n-1}{\sigma-1}+m^2   \ \ .
\ee
 This amplitude takes the form
\bea
&{}&A(s,t)=\nonumber\\
&{}&N\prod_{r=0}^\infty
 \frac{\Bigl((\sigma-1)(s-m^2)+1\Bigr)
\Bigl((\sigma-1)(t-m^2)+1)\Bigr)-\sigma^r}
{\Bigl((\sigma-1)(s-m^2)+1-\sigma^r\Bigr)
\Bigl((\sigma-1)(t-m^2)+1-\sigma^r\Bigr
 ) }\ \ ,
\label{amp}
\eea
where N is a (possibly infinite) normalisation constant.
In a theory with centre of mass momentum $p$, we have
\be
t=-{2p^2}(1-\cos\theta)=-(\frac{s}{2}-2m^2)(1-\cos\theta)\ \ .
\label{leg}
\ee

We can then write a simple procedure to calculate $R(j,k)$, the
coefficient of $P_k(\cos\theta)$ once the residue of
the pole at $s=\lambda_j$ has been expanded in orthogonal Legendre
polynomials. A negative value of $R(j,k)$, the
square of the coupling between the two external scalars and the exchanged
particle of mass square $\lambda_j$ and spin $k$, is the
signature of an exchanged ghost.

A crucial formula \cite{Sneddon}
\be
\int_{-1}^1(1-x)^mP_k(x)dx=\cases{(-1)^k{2^{m+1}(m!)^2\over(m-k)!(m+k+1)!}
&\quad\quad\quad $ m\geq k,$\cr
0&\quad\quad\quad$ m<k.$\cr}
\ee
enables us to project out the  $k$th partial wave of the residue at the pole
$s=\lambda_j$ (see (\ref{spectrum}))
\be
\lambda_j=l(j)+m^2.
\ee
This residue is, as we know, a polynomial of maximum degree
$j$ in $t$. One can then write easily a simple symbolic procedure
to calculate $R(j,k)$.

Using this procedure, we have computed analytically the $R(j,k)$
for quite a number of low values of $j$ and $k$. In particular, it is
possible to deduce
a general formula for cases $R(j,j)$ and $R(j,j-1)$ as follows;
\bea\label{Rjj}
R(j,j&)&=\frac{(j!)^2}{g(j)(2j)!}
(l(j)-3m^2)^j,\nonumber\\
R(j,j-1&)&=\frac{((j-1)!)^2}{2g(j)(2j-2)!}(l(j)-3m^2)^{j-1}
\times\\
&{}&(jm^2-jl(j)+\frac{2}
{\sigma^j}\frac{\partial l(j+1)}{\partial\sigma}),\nonumber
\eea
where as usual $l(j)$ is given by (\ref{spectrum}) and
\be
g(j) =\frac{ (\sigma-1)(\sigma^2-1))(\sigma^3-1)
\cdots(\sigma^j-1)}{(\sigma-1)^j
\sigma^{\frac{j(j+1)}{2}}}.
\label{next}
\ee

We have explored systematically these analytical $R(j,k)$ for all values
of $k$
within a large range of $j$, looking for patterns inside the interesting
domain $D$ given by $-1\leq m^2\leq 0$ and $\sigma>0$. For later reference let
us take
$m^2$ as the horizontal axis and $\sigma$ as the vertical axis. The domain $D$
is then a semi-infinite vertical strip.

Using the explicit general forms (\ref{Rjj}) and the set of examples we
have shown that within the domain $D$ the behaviour of $R(j,k)$
for $j+k$ even is different from the behaviour for $j+k$ odd. Indeed~:
\begin{itemize}
\item[1)] For even $j+k$, $R(j,k)$ is always positive. The curve given by
$R(j,k)=0$ does not cross the domain $D$.
\item[2)] For odd $j+k$, $R(j,k)=0$ is a curve which crosses the domain
$D$ once only and divides it into two parts. In the low $\sigma$ part $R(j,k)$
is positive while in the higher $\sigma$ part it is negative. The curve
obviously depends on the values $j$ and $k$ but, and this is remarkable,
it is always confined to a small region in the domain.
\end{itemize}

We now discuss the curve $R(j,k)=0$ in a more precise way when $j+k$ is
negative.

First let us examine the crucial $R(1,0)=0$ curve. Its
analytical form is given by
\be
m^2=1-\frac{2}{\sigma}
\label{R10}
\ee
i.e. a hyperbola passing through the points $m^2=-1,\sigma=1$ and
$m^2=0,\sigma=2$, its concavity turned upwards and with a positive
slope inside $D$

All the other curves $R(j,k)=0$ with $j+k$ odd are like the $R(1,0)=0$
curve within the domain $D$. They pass through the point
$m^2=-1,\sigma=1$. Their concavity is upwards and they have a positive slope
inside $D$, though they are no longer hyperbolas but higher degree polynomials.
Finally the intersection point in the vertical axis $m^2=0$ has a
coordinate $\sigma=\sigma_{jk}$ and
\be
1<\sigma_{jk}\leq 2.
\label{bjk}
\ee
Numerical estimates show that, when $j$ increases, $\sigma_{jk}$, while
decreasing, converges to 1. We here quote a few values
$\sigma_{10}=2,\ \sigma_{21}=1.246979,
\ \sigma_{32}=1.113625,\ \sigma_{30}=1.111834,\ \sigma_{43}=1.065670,
\ \sigma_{41}=1.064472,\ \sigma_{54}=1.042874,
\ \sigma_{52}=1.042066,\ \sigma_{50}=1.041721.
$

 From these considerations we can draw the following conclusions which were
also tested by a numerical exploration for much higher values of $j$ and
$k$ and can easily seen to be correct for analytical $R(j,j-1)$ given above in
(\ref{Rjj}).

\begin{itemize}
\item[1)] for $-1\ \leq m^2\ \leq 0$ and $0<\ \sigma\ \leq 1$
all the $R(j,k)$ are positive and hence the 4 point
amplitude is ghost free.

\item[2)] If the point $m^2,\sigma$ is above the curve (\ref{R10}), then
the amplitude is
not ghost free, but systematically $R(j,k)>0$ if $j+k$ is even and
$R(j,k)<0$ if $j+k$ is odd. In this case, which would correspond to an
exponentially rising $\lambda_n$ since $\sigma>1$, it is not impossible
that a mechanism
might be devised to reinterpret those wrong sign poles, since the pattern of
their occurrence is so simple.

\item[3)] In the intermediate range i.e. when $\sigma$ lies between the line
$\sigma=1$
and the curve (\ref{R10}),
 the results
have a pattern in $R(j,k)$ which varies with $\sigma$ and $m^2$. For all
even $j+k$ the sign is systematically positive. For odd $j+k$,
on a given point $m^2,\sigma$ in the intermediate range, some
$R$'s at low values of $j$
and $k$ are positive while they are all negative at higher
values of $j$ and $k$.

\item[4)] If the patterns we have found are correct (and we
believe they are) they show that $\sigma=1$, the Veneziano case,
is exactly the borderline case of the region where the
interpretation of the amplitude in terms of exchanged particles
of definite spins holds good. In this limit, when the mass $m^2$
approaches the tachyon
value -1, the intermediate range shrinks to zero. At the
transition point to the
alternating pattern, which occurs at $\sigma=1$, all the $R(j,k)$
are equal to 0 for odd $j+k$.
This is the statement that the odd daughters decouple.
\end{itemize}

The case envisaged by Baker and Coon and also Cremmer and Nuyts corresponds to
the regime where $\sigma>1$ and thus there is ghost behaviour.

The
ghost free case, $0\leq\sigma\leq 1$, as we have remarked in the introduction
implies that the spectrum in mass(squared) approaches a limit point
\be
\displaystyle{\frac{\sigma}{1-\sigma}+m^2}.
\ee
We  stress that
when $\sigma$ is very close to 1 (i.e. when we are close to the
linear case $\sigma=1$), the limit point may be made arbitrarily
high (tentatively the Planck mass~?) and that the low lying resonances
lie on a trajectory experimentally indistinguishable from a linear
trajectory. This would be appropriate for an interpretation in terms of
hadronic
amplitudes.

The existence of an accumulation point in the spectrum obviously implies a
breakdown of
the validity of the expression for the scattering amplitude at this $s$ value.
However, this may be reconcilable with current attempts to introduce a
fundamental length via string theories. (See, for example \cite{asp}).

\section{Conclusion}

In this paper we have re-opened
some old questions concerning the nature of
the dual resonance model and its
realisation as a theory of first quantized strings. The principal new result is
the conjecture, based upon extensive empirical observation, that the Veneziano
 amplitude admits a deformation
which preserves the properties of duality and in particular, is ghost free.
The related trajectory has an accumulation point which can be
made to occur at an arbitrarily high value of the energy.
This strongly suggests the existence of a deformed string theory to
give an interpretation to such amplitudes.
 Is there an operator formalism which
reproduces these results, involving, maybe, quantum creation and
annihilation operators~?

 We have also shown that the M\"obius trajectory allows   the
construction of a dual 4-point amplitude describing the exchange
of an infinite number of particles or resonances of maximum spin
$n$ increasing monotonically along the trajectory. It is a more general
solution than the logarithmic trajectories with exponential spectrum, since
it depends upon the four parameters which may be taken to be the masses of the
first four poles  and some of them can even be chosen
as complex.
A secondary series of questions is concerned with the amplitudes for the
 M\"obius trajectory.
Is there anything corresponding to an integral representation for our M\"obius
 amplitudes~?
There are two additional parameters in our 4-point amplitude;
$\hat\alpha_0$ and
$\hat\alpha_1$. Can we exploit this freedom and
extend to a viable 5-point function~?

In summary, we believe
that the miraculous way in which the 4 parameter spectrum works
for the M\"obius trajectory and the absence of ghosts in certain
ranges of the amplitudes for the logarithmic trajectories with an
accumulation point
suggest a new avenue of exploration in string theory.
\vskip 10pt
\centerline{\bf Acknowledgement}
\vskip 10pt
J. Nuyts acknowledges partial support for this work from the European Community
``Human Capital and Mobility'' grant ERB-CHR-XCT 920069.

\hfill\eject


\begin{thebibliography}{**}
\bibitem{ven} G. Veneziano,
             Nuovo Cimento, {\bf{57A}} (1968) 190.
\bibitem{fub} S. Fubini, D. Gordon and G. Veneziano,
             Phys. Letters, {\bf{29B}} (1969) 679.
\bibitem{bc} D.D. Coon and M. Baker,
             Phys. Rev., {\bf{D2}} (1970) 2349.
\bibitem{cn} E. Cremmer and J. Nuyts,
             Nuc. Phys., {\bf{B26}} (1971) 151.
\bibitem{go} J.M. Golden, Lett.
             Nuovo Cim., (Ser.2) {\bf{1}} (1971) 893.
\bibitem{ma} S. Machida,
             Prog. Theor. Phys., {\bf{47}} (1972) 2015.
\bibitem{ro} L.J. Romans,
             in {\it Proceedings of the High Energy Physics and Cosmology"
             Conference, Trieste,} (1989),
             eds. J.C. Pati et al., World Scientific (1990).
\bibitem{coon3} D. Coon, S. Yu and M. Baker,
             Phys. Rev., {\bf{D5}} (1972) 1429.
\bibitem{masud} M. Chiachian, J.F. Gomes and P. Kulish,
             Phys. Letters, {\bf{B311}} (1993) 93.
\bibitem{coon2} D. Coon,
              Phys. Letters, {\bf{29B}} (1969) 669.
\bibitem{asp} P. Aspinwall,
             {\it  Minimum Distances in Non-Trivial String Target Spaces},
              preprint IASSNS-HEP-94/19, (1994).
\bibitem{br} K. Bardakci and H. Ruegg,
              Phys. Letters, {\bf{28B}} (1968) 342.
\bibitem{god} P. Goddard and C. Thorn,
              Phys. Letters, {\bf{40B}} (1972) 235.
\bibitem{brow} R.C. Brower,
              Phys. Rev., {\bf{D6}} (1972) 1655.
\bibitem{fai} D.B. Fairlie and J. Nuyts,
              Jour. Phys., {\bf{A24}} (1991) L1001.
\bibitem{jack} F.N. Jackson, Quart. J. Pure \& Appl. Math. {\bf 41} (1910) 193.
\bibitem{Sneddon} I.M. Sneddon,
              {\it Special Functions of Physics and Chemistry}
              Oliver and Boyd, (1956) Edinburgh.

\end{thebibliography}
\end{document}